**Title: On the prevalence of small-scale twist in the solar chromosphere and transition region**

**Authors:** B. De Pontieu[1,2], L. Rouppe van der Voort[2], S.W. McIntosh[3], T.M.D. Pereira[2], M. Carlsson[2], V. Hansteen[2], H. Skogsrud[2], J. Lemen[1], A. Title[1], P. Boerner[1], N. Hurlburt[1], T.D. Tarbell[1], J.P. Wuelser[1], E.E. De Luca[4], L. Golub[4], S. McKillop[4], K. Reeves[4], S. Saar[4], P. Testa[4], H. Tian[4], C. Kankelborg[5], S. Jaeggli[5], L. Kleint[6], J. Martinez-Sykora[6]

**Affiliations:**

[1] Lockheed Martin Solar and Astrophysics Laboratory, 3251 Hanover St., Org. A021S, Bldg.252, Palo Alto, CA, 94304, USA.

[2] Institute of Theoretical Astrophysics, University of Oslo, Post Office Box 1029, Blindern, N-0315, Oslo, Norway.

[3] High Altitude Observatory, National Center for Atmospheric Research, Post Office Box 3000, Boulder, CO 80307, USA.

[4] Harvard-Smithsonian Center for Astrophysics, 60 Garden St., Cambridge, MA 02138, USA.

[5] Department of Physics, Montana State University, Bozeman, P.O. Box 173840, Bozeman MT 59717, USA.

[6] Bay Area Environmental Research Institute 596 1st St West, Sonoma, CA, 95476 USA.

*Correspondence to: bdp@lmsal.com

**Abstract:** The solar chromosphere and transition region (TR) form an interface between the Sun's surface and its hot outer atmosphere. Here most of the non-thermal energy that powers the solar atmosphere is transformed into heat, although the detailed mechanism remains elusive. High-resolution (0.33-arcsec) observations with NASA's Interface Region Imaging Spectrograph (IRIS) reveal a chromosphere and TR that are replete with twist or torsional motions on sub-arcsecond scales, occurring in active regions, quiet Sun regions, and coronal holes alike. We coordinated observations with the Swedish 1-m Solar Telescope (SST) to quantify these twisting motions and their association with rapid heating to at least TR temperatures. This view of the interface region provides insight into what heats the low solar atmosphere.

**One Sentence Summary:** Observations with the Interface Region Imaging Spectrograph (IRIS) reveal that the solar chromosphere and transition region are replete with small-scale twisting motions that are associated with rapid heating to transition region temperatures.

**Main Text:** The physical mechanism that is predominantly responsible for the heating of the solar outer atmosphere remains unknown. A variety of mechanisms are still under investigation, the most important ones being dissipation of magnetic waves *(1-4)* or braiding and reconnection of magnetic fields and subsequent energy release *(5-7)*. One mechanism that has recently received attention is the effect of vorticity *(8)*, assumed to be generated by magneto-convective motions at or below the surface, within the outer atmosphere. Theoretical models predict that vorticity may be a signature of reconnection or magnetic waves, or may be associated with strong electrical currents, all of which can lead to significant heating *(8)*. However, observational support in the solar atmosphere for vortical motions and associated heating has been (i) limited to the chromosphere, *i.e.,* lacking a heating component *(9-10)*, (ii) limited to quiet Sun regions (i.e., away from active regions) *(10)*, or (iii) isolated in incidence and on larger spatial scales *(11,12)* than predicted by models *(13,14)*.

We report observations of twisting or torsional motions that are much more prevalent than previously reported *(12)*, permeate both the chromosphere and TR of the Sun, and sometimes appear to be associated with vigorous heating of chromospheric plasma. We exploit the multi-thermal chromospheric (<20,000 K) and TR (20-80,000 K) coverage of the high spatial (0.33 arcseconds) and spectral (3 km/s per pixel) resolution near-ultraviolet (NUV) and far-ultraviolet (FUV) spectra and slit-jaw images taken with the 20-cm telescope onboard the Interface Region Imaging Spectrograph (IRIS, *15*), which was launched in June 2013. These data reveal the strong torsional motions (10-20 km/s) that imply significant twist in the chromospheric and TR magnetic field.

We analyze spectroheliograms, i.e., raster scans in which the IRIS slit is scanned across the solar surface to build up a three-dimensional view (two spatial, one spectral dimension) in the Mg II h 2803 Å and Si IV 1402 Å spectral lines, which are formed, respectively, in the chromosphere (5,000-15,000 K) and TR (80,000 K, under ionization equilibrium conditions) *(12)*. From that data, we constructed Dopplergrams *(12)*, in which we subtracted the intensities in the blueshifted and redshifted wings of a spectral line (at fixed offset velocities from the rest wavelength, e.g., ±30 km/s). The Dopplergrams reveal a large number of elongated loop-like structures that contain regions of strongly blue- and redshifted plasma that are in close proximity to one another and that are part of the same dynamic structure (Figs. 1, 2, S7, S8 and Movies S1-S20). Most of the plasma imaged in these spectral lines occurs at heights where the magnetic field dominates the plasma dynamics (plasma $\beta \ll 1$) *(15, 16-19)*. The elongated structures thus most likely trace the magnetic field, and the observed plasma motions are compatible with torsional flows, i.e., flows along twisted magnetic field lines – alternative interpretations have been considered, but deemed less likely *(12)*. Fig. 1 shows the ubiquity of these twisted features in active regions and quiet Sun on the disk and at the limb. The prevalence of these twisted features is actually higher than observed here for several reasons *(12)*. The torsional motions are only visible if the viewing angle is significantly inclined with respect to the magnetic field direction, which is not the case over the whole field of view. In addition, both field-aligned flows and the significant swaying motions in the solar atmosphere *(2, 4, 10, 20)* render it less likely to see the "ideal" blue/red appearance of a twisted feature.

Despite these limitations, it is clear that twist occurs throughout much of the field of view within a variety of structures (Figs. 1, 2, S7, S8 and Movies S5-S21). In quiet Sun regions, many of these features appear to be spicules *(20)*, rapidly evolving jets that appear to propel plasma upwards and on which torsional motions have been reported before in chromospheric lines *(10,21)*. However, the IRIS observations now reveal the heating associated with the twisting motions (Fig. 1): the hotter Si IV emission occurs towards the tops of the spicules, both at the limb and more clearly on the disk (where there is less superposition). Moreover, IRIS now reveals prevalent twist also in active region loops that appear to be low-lying and highly dynamic, with both chromospheric and TR plasma involved in the torsional motions (Figure 1B, Movies S5-S7, S21). We also see twist in some of the small-scale loops that appear to continuously form in quiet Sun regions *(22)* and that are visible at the limb.

The amplitude of the torsional motions was estimated from a detailed study of the spectra across the twisted features. Several examples shown in Fig. 2 (and S8) indicate that velocities of order 10-30 km/s (over a cross-field distance of 1 arcsec or less) appear to be typical with one side of the feature more redshifted, and the other side more blueshifted, leading to a tilted appearance in wavelength-space plots. Such tilted features are ubiquitous, especially towards the limb where the viewing angle combined with the mostly radial magnetic field provide optimal visibility. There are indications that there may even be more locations with significant twist that are not resolved by IRIS. For example, many locations show spectra that do not appear to be tilted, but are instead very broad. Such broad profiles could also be caused by bidirectional field-aligned flows or other sources of non-thermal broadening such as small-scale turbulence. Our observations suggest that at least a significant subset of these broad profiles is likely caused by torsional motions *(12)*.

The temporal and thermal evolution of these twisted features is brought to light even more clearly with simultaneous IRIS slit-jaw images *(12)* and rapid scans in the chromospheric Hα spectral line (Figure 3) taken with the CRISP Fabry-Pérot interferometer *(23)* at the SST *(24)*. Chromospheric Dopplergrams show that the twisted features are highly dynamic (Figure 3 and movies S5-S19) with evidence of rapid propagation of the torsional motions along the features, typically 30-100 km/s (Fig. 4, S6). This speed is compatible with the value expected for Alfvén waves in the upper chromosphere and low TR. Typical timescales for the torsional motions are of the order of 1 minute (Fig. 2C, 2D) with short-lived excursions in the far blue and red wings of the Mg II h 2803Å line predominant in regions where the line-of-sight is more perpendicular to the magnetic field. The torsional timescales stand in sharp contrast to the longer timescales (several minutes) and smaller line-of-sight velocities in plage regions (Fig. 2B) where the line-of-sight is more aligned with the magnetic field and the line profiles are dominated by magneto-acoustic shocks *(25)*.

Slit-jaw images (SJI) taken with IRIS (Fig. 3 and Movies S8-S10) also show clearly how the propagating torsional motions are often associated with bright, highly dynamic linear features (C II 1335 Å and Si IV 1402 Å); this emission indicates heating to at least TR temperatures (20,000-80,000 K). In quiet Sun regions, these SJI features appear to be the TR counterparts of so-called rapid-blueshifted and rapid redshifted events in the chromosphere, the disk counterparts of spicules *(26)*. Many of the TR features IRIS observes in quiet Sun regions are associated with twisted features, indicating that the heating we observe is substantial. Some of these events are also visible in coronal passbands (movie S7), although it is not yet clear how much plasma is heated to coronal temperatures in association with these heating events *(12)*.

Our observations of twist that permeates the chromosphere and TR of the Sun dramatically expand on a picture that has recently emerged that draws together disparate observations of twist in the solar atmosphere: in macrospicules *(27,28)*, explosive events *(28)*, spicules *(10,26)*, and so-called swirls *(11,12)*. The powerful combination of SST and IRIS observations indicate that, as higher spatial resolutions become available, the apparent prevalence of twist drastically increases. In addition, the unique TR coverage of IRIS shows that the presence of twist is not limited to the chromosphere, but extends into higher temperature regimes, and indicates that significant heating often occurs in twisted features.

The occurrence of this twist and the associated heating likely has several causes. It seems possible that the strong photospheric vortical flows that have been observed *(29)* and that occur in advanced numerical models *(8)* play a significant role. In such models the vortical flows are often associated with strong currents, and it thus seems plausible that significant heating could result. Some of the observed twist, e.g., in the small-scale loops observed with IRIS *(22)*, likely originates from twist in flux tubes that emerge into the atmosphere. Although more advanced modeling and further coronal observations are required, the observed heating is also compatible with recent numerical models that predict significant chromospheric and coronal heating from dissipation of torsional Alfvén waves generated by very small-scale photospheric vortices *(13,14)*. The prevalent twist and associated heating are also compatible with dissipation of torsional modes that arise from resonant absorption of swaying motions on flux-tube like features in the atmosphere *(30)*. Finally, torsional motions and heating could also be expected if they result from the reconnection of field lines, e.g., as a result of flux emergence *(16)*. It is thus clear that detailed studies of the origin of twist and extent of the heating will open a new window on several proposed heating mechanisms for the lower solar atmosphere. Such studies should also help elucidate the impact of the ubiquitous propagation of small-scale twist on the helicity budget of the solar atmosphere, which may play a role in solar eruptions.

Acknowledgments: We thank A. Ortiz, E. Scullion, A. Sainz-Dalda for assistance with SST observations. IRIS is a NASA Small Explorer developed and operated by LMSAL with mission operations executed at NASA Ames Research center and major contributions to downlink communications funded by the Norwegian Space Center through an ESA PRODEX contract. IRIS data can be downloaded from iris.lmsal.com. This work is supported by NASA contract NNG09FA40C (IRIS), the Lockheed Martin Independent Research Program and the ERC grant No. 291058. The Swedish 1-m Solar Telescope is operated on the island of La Palma by the Institute for Solar Physics of Stockholm University in the Spanish Observatorio del Roque de los Muchachos of the Instituto de Astrofísica de Canarias.

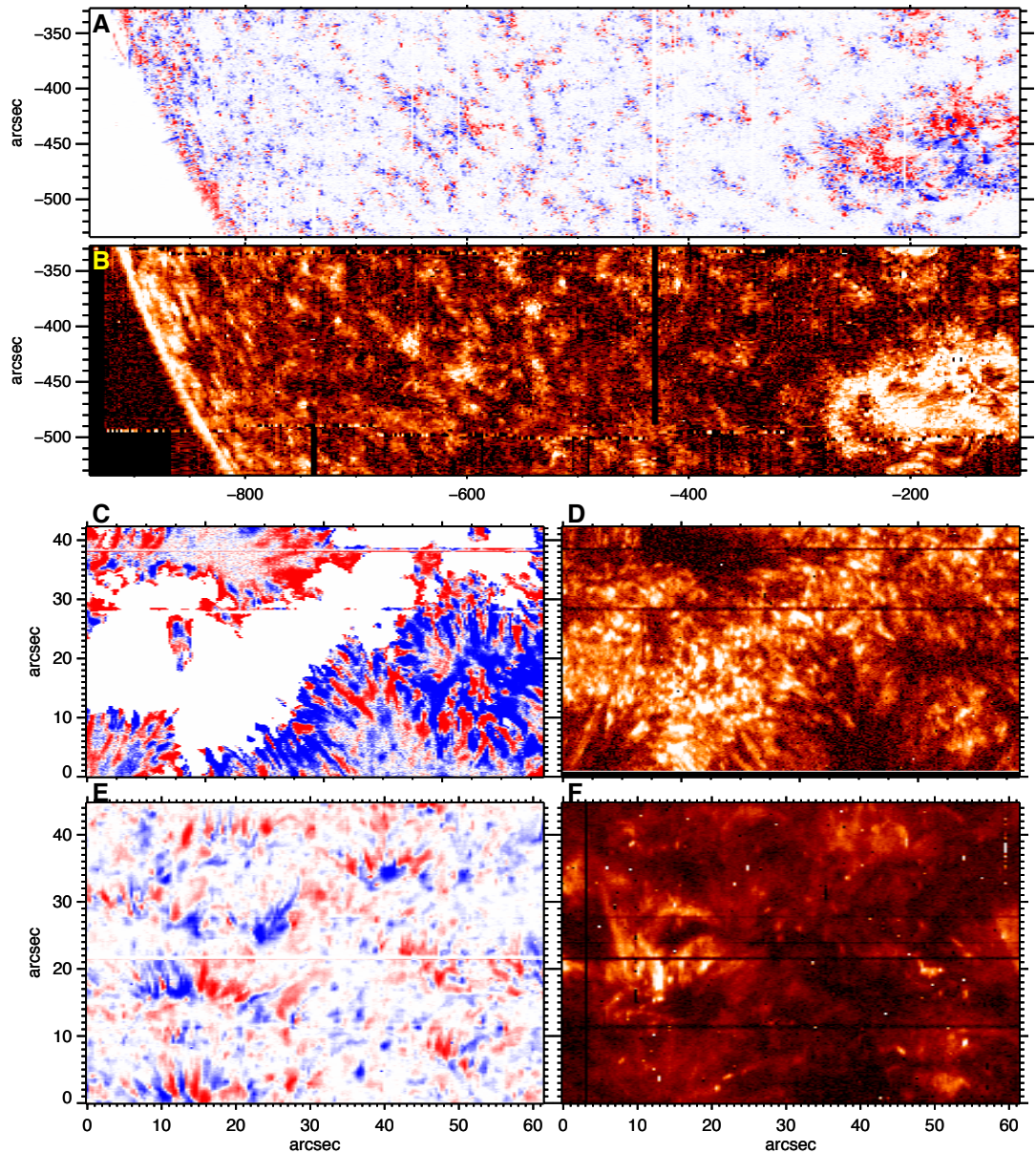

**Fig. 1.** Prevalence of twist in quiet Sun and active regions. Dopplergrams (panels A, C, E) of the chromospheric (~10,000K) Mg II h 2803Å line (at 30 km/s from line center) show a multitude of elongated features in which strongly red- and blueshifted features are parallel and adjacent to each other. These features illustrate that twist is predominant: at the solar limb associated in so-called spicules (A, E), but also in active regions (panel C, and around -250",-475" in panel A), both regions where the line-of-sight is likely more perpendicular to the local magnetic field. Twist is often associated with significant brightening in the transition region (80,000 K) as illustrated with the Si IV 1403Å integrated brightness map in panels B, D and F. When looking straight down plage regions the line-of-sight is aligned with the magnetic field, severely reducing or eliminating the visibility of twist. This is why the plage regions have been removed from the active region Dopplergram. See Fig. S7 for a larger field-of-view.

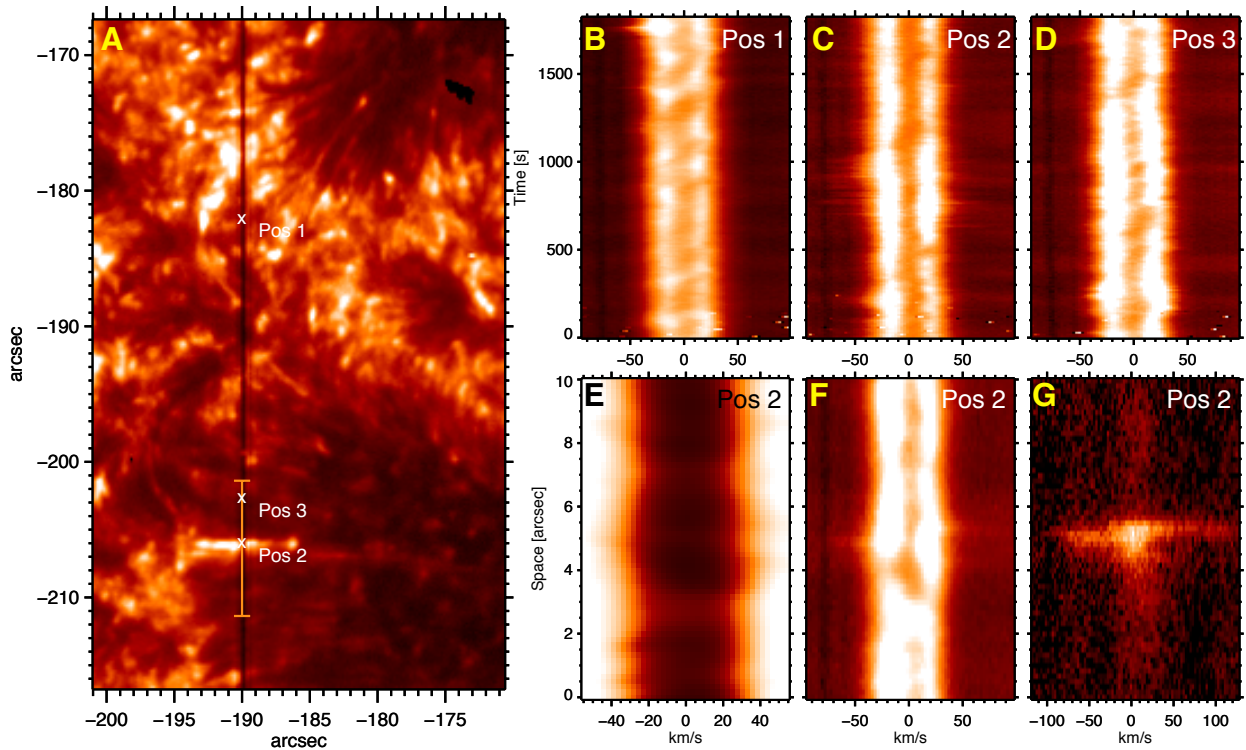

**Fig. 2.** Spatio-temporal properties of twist. Rapidly evolving twisting motions are apparent as short-lived, bright features in the blue and red wings (e.g., around ±50 km/s) of the chromospheric Mg II h 2803Å spectral line (panels C, D) in regions of inclined field (positions 2, 3 shown in panel A, Si IV 1403Å slit-jaw image). These motions are in contrast to the acoustic shock-dominated spectral profiles (panel B) in position 1 in plage (where field and line-of-sight are more aligned, preventing visibility of twist) that evolve on time-scales of several minutes *(25)*. The spatial pattern of the bright Si IV feature around -190",-207" (position 2 in panel A) is associated with short-lived twisting motions that are visible in Si IV (80,000K, panel G), Mg II h (10,000K, panel F) and faintly in H-alpha (<10,000K, panel E, see movies S5-S7). Velocities of order 50 km/s are reached in this twisting feature. Typical velocities are lower (10-30 km/s) with visibility in these various passbands variable, Mg II h showing excellent visibility in most cases. See Supplementary Materials for other examples (Figs. S7, S8) and movies S5-S19, S21.

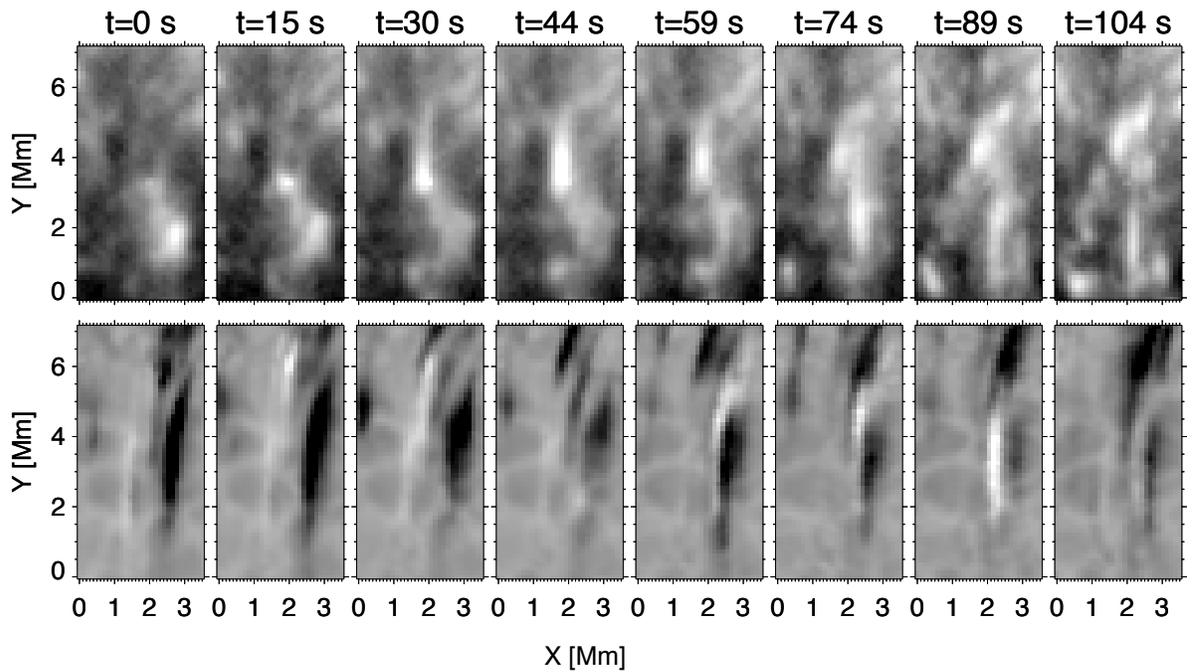

**Fig. 3.** Temporal evolution of twist and associated heating. SST Hα Dopplergrams at ±46 km/s (bottom row, black is blueshifted, white is redshifted) show how quickly chromospheric twist propagates along elongated features on timescales of less than 1 minute. Several of these twisted features are associated with TR signals (top row) as observed with the IRIS slit-jaw images that are dominated by Si IV lines (~80,000 K). Movies S8-S10 provide further examples of the dynamic nature of the twist and associated heating.

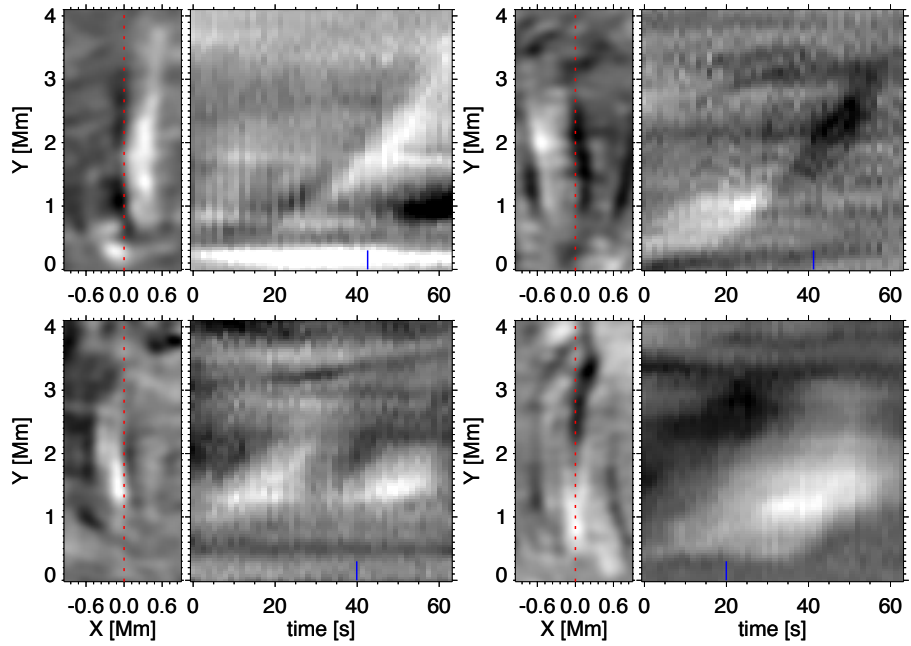

**Fig. 4.** Propagation speed of twisting motions near the limb. Chromospheric Ca II 8542Å Dopplergrams at ±25 km/s of twisted features (left side of panels A, B, C and D) from high cadence observations at the SST show rapid propagation (~30-100 km/s). These speeds are derived from time-distance plots (right side of panels) of the Doppler signal along the long axis of the features (red dashed line) and are consistent with Alfvén speeds at chromospheric heights..

Supplementary Materials for

# On the Prevalence of Small-Scale Twist in the Solar Chromosphere and Transition Region

**Authors:** B. De Pontieu[1,2], L. Rouppe van der Voort[2], S.W. McIntosh[3], T.M.D. Pereira[2], M. Carlsson[2], V. Hansteen[2], H. Skogsrud[2], J. Lemen[1], A. Title[1], P. Boerner[1], N. Hurlburt[1], T.D. Tarbell[1], J.P. Wuelser[1], E.E. De Luca[4], L. Golub[4], S. McKillop[4], K. Reeves[4], S. Saar[4], P. Testa[4], H. Tian[4], C. Kankelborg[5], S. Jaeggli[5], L. Kleint[6], J. Martinez-Sykora[6]

**Affiliations:**

[1]Lockheed Martin Solar and Astrophysics Laboratory, 3251 Hanover st., Org. A021S, Bldg.252, Palo Alto, CA, 94304, USA.

[2] Institute of Theoretical Astrophysics, University of Oslo, Post Office Box 1029, Blindern, N-0315, Oslo, Norway.

[3] High Altitude Observatory, National Center for Atmospheric Research, Post Office Box 3000, Boulder, CO 80307, USA.

[4]Harvard-Smithsonian Center for Astrophysics, 60 Garden st., Cambridge, MA 02138, USA.

[5] Department of Physics, Montana State University, Bozeman, P.O. Box 173840, Bozeman MT 59717, USA.

[6] Bay Area Environmental Research Institute 596 1st St West, Sonoma, CA, 95476 USA.

*Correspondence to: bdp@lmsal.com

**This PDF file includes:**

    Materials and Methods
    SupplementaryText
    Figs. S1 to S8
    Tables S1
    Captions for Movies S1 to S22

**Other Suplementary Materials for this manuscript includes the following:**

    Movies S1 to S22

# 1. Materials and Methods

1.1 Observations

1.1.1 IRIS Observations

We use several different IRIS datasets, listed in table S1. All of the IRIS data were calibrated to level 2, i.e., including dark current, flat-field and geometric correction, as well as co-alignment between various channels in the far-ultraviolet (FUV) and near-ultraviolet (NUV) passbands *(15)*. We also corrected for the spectral drift associated with the spacecraft's orbital velocity (relative to the Sun) and the thermal drift caused by thermal variations over the course of one orbit. The latter was done by subtracting the long-term trend of the photospheric Ni I 2799.47Å line from our observed wavelength positions. The raster scans are created by scanning the IRIS slit across the solar disk at either dense (0.35"), sparse (1") or coarse (2") steps. At each location of the raster scan spectra are obtained during an exposure time (Table S1) that is the same for all steps in the raster scan. The total duration of the raster scan is given by the exposure time (and overhead) at each location multiplied by the number of steps in the raster scan. The slit-jaw images (SJI) were corrected for dark-current, flat-field, internal co-alignment drifts.

| Date | Type | FOV [arcsec] | Step, Pixel Size [arcsec] | Exposure time [seconds] | Raster Cadence [seconds] | Pointing | Figure/ Movie |
|---|---|---|---|---|---|---|---|
| 10-Sep-2013 08:09-11:09 | Two-step sparse raster | 1x50 | 1, 0.166 | 4 | 10 | AR 11838 | 2a-d, f, g, Movies S5-S7 |
| 13-Sep-2013 08:17-14:54 | 4-step dense raster | 1x50 | 0.35, 0.166 | 4 | 15 | Quiet Sun disk center | S4, Movies S13-S14 |
| 22-Sep-2013 07:34-11:04 | Medium sit-and-stare | 0.33x61 | 0, 0.166 | 4 | 5 | Coronal Hole, (538", 283") | S3, Movies S11-S12 |
| 23-Sep-2013 07:09-12:05 | Medium sit-and-stare | 0.33x61 | 0, 0.166 | 4 | 5 | Quiet Sun disk center | 3, Movies S8-S10 |
| 27-Sep-2013 06:24-06:44 | Dense raster scan | 140x175 | 0.35, 0.166 | 2 | 1166 | AR 11850 | 1c, 1d, S5 |
| 9-Oct-2013 13:10-14:09 | Dense raster scan | 140x175 | 0.35, 0.166 | 8 | 3596 | North Pole | 1e,1f, Movies S3-S4 |
| 13-Oct-2013 04:29-19:38 | Coarse raster scan | 124x175 | 2, 0.166 | 2 | 170 | Full-disk Mosaic | 1a, 1b, Movies S1-S2 |

Table S1: Properties of IRIS datasets used in this paper.

## 1.1.2 IRIS Passbands

IRIS obtains spectra in two FUV passbands (FUV1: 1332-1358Å, FUV2: 1389-1407Å) and one NUV passband (NUV: 2783-2834Å). These spectral passbands are sampled with 12.98 mÅ, 12.72 mÅ and 25.46 mÅ for, respectively, FUV 1, FUV 2 and NUV. IRIS obtains high-resolution (0.33 arcsec in FUV, 0.4 arcsec in NUV) spectra in C II 1334Å and C II 1335Å (both formed in the chromosphere and low transition region, 20,000-30,000K), Si IV 1394 Å and Si IV 1403Å (transition region, 80,000 K) and Mg II k 2796Å and Mg II h 2803Å (chromospheric, from 5,000-15,000 K). The estimated temperatures are assuming ionization equilibrium *(15)*. The spectral sampling corresponds to a velocity sampling of about 2.7 km/s. The spatial resolution is 0.33 arcseconds along the slit, whereas the raster step size and Nyquist criterion determine the spatial resolution in the direction perpendicular to the slit: dense rasters have a spatial resolution across the slit of twice that of the step size (0.35 arcseconds), i.e., 0.7 arcseconds.

We use slit-jaw images in the 1330Å, 1400Å and 2796Å passbands. These contain emission from the C II 1334/1335Å lines (chromosphere and low transition region, SJI 1330), Si IV 1394/1403Å lines (transition region, SJI 1400) and Mg II k 2796Å (chromosphere, SJI 2796). All of these SJI channels have a broad passband that also includes continuum formed in the upper photosphere (SJI 1330, SJI 1400) as well as photospheric wing contributions (SJI 2796). We can distinguish between photospheric and transition region contributions by comparison with the clean spectra in the IRIS spectrograph (when observing close to the slit), or through their temporal evolution and morphology. Photospheric contributions are typically either associated with the gradual evolution of so-called granules (magneto-convective cells on the solar surface) on timescales of minutes, or extremely short-lived bright points associated with the propagation of magneto-acoustic shocks into the atmosphere *(31, 32)*. The signals we study here in our 1330 and 1400 SJI images in association with the chromospheric twist are loop-like or jet-like linear features that are clearly not photospheric in origin but formed in the transition region lines C II 1334Å, C II 1335Å or Si IV 1394Å and Si IV 1403Å.

## 1.1.3 IRIS Full Disk Mosaic

The IRIS full-disk observation shown in Figure 1a, 1b was taken on 13-October-2013 from 04:29 - 19:38 UT. The observation comprised 165 individual pointings of the spacecraft/telescope. The field of view of each pointing was 124" x 175" with IRIS sampling the region in 3 minutes with 2" resolution in x and 0.33" resolution in y, full spectral resolution and an exposure time of 2s. Because of the slight uncertainty of order 5" in pointings, some areas of the Sun were not covered by the full-disk mosaic. These regions do not contain any valid data and are visible as vertical stripes that are black in intensity maps, and bright white in Dopplergrams (Fig 1a-1b, Movies S1-S2).

### 1.1.4 Swedish 1m Solar Telescope

We use observations from the CRISP tunable filter instrument *(24)* at the Swedish 1-m Solar Telescope *(23)* on the island of La Palma, Spain. CRISP is a Fabry-Perot interferometer that is capable of fast wavelength switching so that densely sampled spectral profiles can be obtained in short time (typically ~4 wavelength positions in less than 1 s) over a field of view of approximately 60 arcsec x 60 arcsec. High spatial resolution near the diffraction limit of the telescope can be achieved with the aid of adaptive optics and post-processing using the Multi-Object Multi-Frame Blind Deconvolution (MOMFBD) image restoration method *(33)*.

The Hα data shown in Figure 2e and Movies S5-7 is from a single line CRISP program sampling the Hα spectral line at 25 line positions with 100 mÅ interval and a cadence of 5.5 s. The Hα data shown in Figures 3, S3, S4, and Movies S8-14 are from a triple line CRISP program acquiring spectral profiles for the chromospheric Ca II 8542Å and Hα lines and photospheric single-wavelength Fe I 6302 -48 mÅ magnetograms at a cadence of 10.8 s. Hα is sampled at 15 wavelength positions with 200 mÅ interval (acquisition time 3.4 s). The data used for Figure 4 and movies S15-19 is from a single line CRISP program sampling the Ca II 8542Å line at 6 wavelength positions (±1100, ±700, ±300 mÅ) at a cadence of 1.33 s.

Seeing variations over the time it takes to complete a spectral scan can introduce local misalignment and compromise the spectral integrity. This is a concern for the analysis of Dopplergrams that are based on the subtraction of sequentially recorded wavelength positions. However, this effect is to a large extent reduced through the use of an independent wide-band channel in the MOMFBD restoration. Residual effects are further attenuated following a different post-restoration method *(34)*.

We note that for Figure 3 the SST data has been interpolated onto the IRIS sjitjaw time grid at an equivalent cadence of slightly less than 15 seconds.

### 1.1.5 Solar Dynamics Observatory

For Movies S5-7 that are supplementary to Figure 2 we include observations from the Atmospheric Imaging Assembly (AIA) *(35)* onboard the Solar Dynamics Observatory (SDO) *(36)*. AIA images for the different channels were calibrated to level 1.5 using the standard aia_prep procedure in SolarSoft. Co-alignment between the AIA channels after calibration was verified and improved by cross-correlation over an area of several arcminutes.

1.2 Methods

1.2.1 Co-alignment

Co-alignment of the SST and IRIS data was done by cross-correlation between CRISP images and IRIS slit-jaw images. The higher resolution CRISP images were scaled down to IRIS resolution. When available, photospheric IRIS 2832 Å slit-jaw images were used for the alignment of far wing Ca II 8542 Å or Hα images that show similar photospheric scene. If the 2832 Å channel was not part of the IRIS program, Mg II k 2796 Å slit-jaw images were used for the alignment of Ca II 8542 Å or Hα wing images. The precision in alignment was found to be at the level of the IRIS pixel scale of 0.166 arcsec. Co-alignment of the AIA and IRIS data was done by cross-correlation between AIA 1600 Å and IRIS Mg II k 2796 Å slit-jaw images. The precision in alignment was found to be at the level of the AIA pixel scale of 0.6 arcsec.

1.2.2 Dopplergrams

In this paper we extensively use so-called Dopplergrams to illustrate the presence on small spatial scales of twisting motions. Dopplergrams exploit the Doppler effect that shifts photons emitted by gas moving towards the observer towards the blue (shorter wavelengths) and photons emitted by gas moving away from the observer toward the red (longer wavelengths). A Dopplergram is formed by subtracting signals in the blue wing of a spectral line from those in the red wing, thus accentuating velocity differences (Fig. S1). Close to the line core of an optically thick line such as Mg II h 2803 Å the Dopplergrams do not necessarily indicate flow patterns since opacity effects are responsible for the general shape of the line close to the core *(17-19)*. However, in the far wings of the Mg II h line the presence of a strong signal cannot be explained by opacity effects, but rather indicates strong flows *(17-18)*.

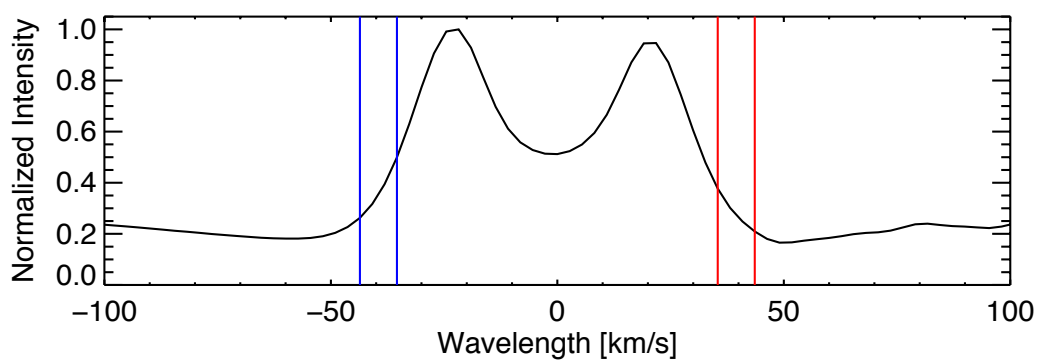

Figure S1: Normalized average Mg II h 2803Å spectrum obtained on 9-Oct-2013 illustrating how Dopplergrams are formed by subtracting signals in the blue wing (between both vertical blue lines on the left) from the red wing (between both vertical red lines on the right).

Since the goal of the Dopplergrams in this paper is to accentuate the presence of red/blue pairs, we have chosen a zero velocity for the Dopplegrams that provides a balanced red/blue appearance at high Dopplergram velocities. The shifts from the "real zero velocity base" are of order 1 km/s, so not significant. This slight deviation from zero km/s is because of the very slight difference between the very far blue and very far red wing of Mg II h. Note that we use Mg II h instead of Mg II k since the latter has a blend in the blue wing.

## 2. Supplementary Text

### 2.1 Comparison with previous observations

Rotational motion has been reported to exist on a variety of spatial scales in the solar atmosphere, but not with the ubiquity, on the small spatial scales or with the TR response we report on here. For example, chromospheric swirls *(37)*, observed in Ca II 8542 Å, were found to display associated signals in SDO/AIA transition region and coronal channels *(11)*. Chromospheric swirls are different in several ways from the linear structures for which we here report ubiquitous torsional motions at sub-arcsecond scale. Swirls occur on larger scales: they consist of rotating arcs or ring fragments and show rotation on larger spatial scale (typical diameters of order 2 arcsec) and larger temporal scale (lifetime typically longer than 10 minutes). Swirls are also nowhere near as ubiquitous as the twist we report on here: swirls are primarily detected in Ca II 8542 Å in quiet regions associated with isolated magnetic bright points away from the network magnetic fields that are dominated by extended linear fibrils in rosetta-like structures. It is particularly in the latter regions we here report the prevalence of smaller-scale twist and torsion and associated heating. Swirls do not typically occur in active region plage and/or quiet Sun/coronal hole network, where we observe most of our twisting features. The observed occurrence rate of swirls is about 20 times less than those of the twisted features reported on here.

Ground-based observations were used to interpret oscillatory behaviour of the linewidth in sparsely sampled, low cadence Hα spectral profiles from one bright point region in quiet Sun as signatures of torsional twist *(9)*. However, these authors do not report on ubiquity or transition region counterparts and do not directly show the twisting motions, reporting instead on line broadening. The timescales reported are much longer (420 seconds) than those of the events reported on in the current paper (<1 min). In addition, the spatial scale of the bright point region is of order 10x10 arcsec, much larger than the small-scale features we report on here.

Explosive events, at spatial scales of several arcseconds, have been proposed to be associated with twisting motions in the transition region *(28,44)*. On larger spatial scales of order 5 arcseconds, macrospicules have been reported to show twisting motions *(27,28,44)*. Finally on spatial scales of 5 arcseconds or more, long-lived (>1 hour) EUV cyclones associated with rotating magnetic field have been observed in the quiet Sun corona *(38)*. Our observations are on much smaller scales, are more ubiquitous and do not correspond to classical explosive events, macrospicules or EUV cyclones.

Chromospheric spicules have been reported to show twisting motions *(10)* but no transition region counterparts or heating were reported, and the observations were limited to quiet Sun regions, so lacked the ubiquity we report on here.

## 2.2 Visibility of twist

Twist on solar features is difficult to detect in general for various reasons:

1. Twist or torsion can only be detected if the twisting motions (which are perpendicular to the main axis of the solar feature) have a significant component along the line-of-sight to the observer. Given the nature of the features we study here (linear jet-like or loop-like features), this means that twist is most easily visible either towards the limb (around quiet Sun or coronal hole network regions) or in inclined field regions in active regions. This is illustrated in Figure 1 where most of the red-blue pairs in Dopplergrams are visible towards the limb, around the quiet Sun network, or at the periphery of active region plage where magnetic field lines are less vertical. The quiet Sun features are associated with so-called spicules or jets *(39)* and short transition region loops *(22)*. The active region features emanate from plage regions and are related to so-called Rapid Blueshifted Excursions (RBE) *(21)*. When looking straight down plage regions the line-of-sight is aligned with the magnetic field, severely reducing or eliminating the visibility of twist. This is why we have removed the plage regions from the active region Dopplergram in Figure 1.

2. If the diameter of the solar feature is at or below the resolving power of the instrument, the torsional motions will be unresolved and lead to non-thermal line broadening. Given the large amount of non-thermal line broadening in optically thin lines in the transition region (10-35 km/s, Fig. S5), our resolved measurements of twist at IRIS resolution of 0.33 arcsec, and the resolved twist at SST resolution of 0.15 arcsec (but below IRIS resolution), it seems likely that some of the non-thermal line broadening observed in TR lines is caused by unresolved twist.

3. To observe twist we typically use Dopplergrams (see SM 1.2.2) with a fixed reference wavelength ("0 km/s"). This means that any other flow or motion that contributes to the line-of-sight velocity such as swaying motions or field-aligned flows will bias the average velocity of the solar feature away from the reference wavelength. This is illustrated in Figure S2 which shows how the velocity across a feature carrying torsional motions of 25 km/s is affected by non-twisting motions projected into the line-of-sight from -30 km/s to +30 km/s. It is clear that as soon as these non-twisting velocities are of order 10-15 km/s the visibility of the twist is severely reduced or eliminated. In the dynamic solar environment such non-twisting motions are known to occur with amplitudes of order 20-60 km/s, so our Dopplergram observations clearly can only show a subset of the twisted features in the solar chromosphere. In particular, spicules are known to show swaying motions of order 10-20 km/s *(2, 43)* as well as field-aligned flows of order 20-60 km/s. The swaying motions are acting in the same direction as the twisting motions, i.e., perpendicular to the long axis of solar features. This means that it is to be expected that twist is often hidden by swaying motions. In addition, while the field-aligned flows are perpendicular to the twisting motions, they are still expected to contribute a significant component to the line-of-sight under a wide variety of viewing angles, further complicating visibility of twist. This is illustrated in movie S20 which shows synthetic Dopplergrams from a Monte Carlo simulation that aims to demonstrate how the twisting and field-aligned motions act together to complicate the interpretation of propagating Alfven waves in our observations. Without accurate knowledge of the magnetic field direction (with respect to the line-of-sight direction) it is difficult to disentangle and determine all three velocity components.

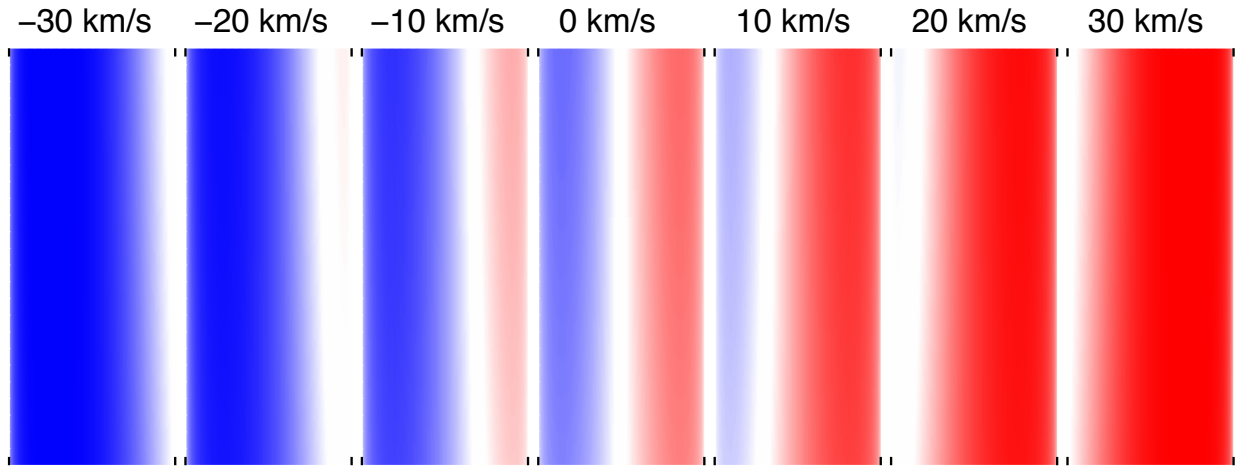

Figure S2: Illustration of limited visibility of twist, for a feature that also includes field-aligned flows and/or swaying motions. A cylindrical structure carrying a torsional Alfven wave with an amplitude of 25 km/s is shown in the middle panel ("0 km/s"). The same structure is shown for various values of the line-of-sight component of any other motions (labels on top from -30 to +30 km/s) that move the whole cylindrical structure towards or away from the observer, e.g., swaying motions or field-aligned flows. Such motions bias the average velocity of the whole feature and thus render the red/blue pattern associated with twist invisible in fixed wavelength Dopplergrams. Note that this Figure includes the effect of the cylindrical shape of the flux tube, unlike the similar Figure in Sekse et al. *(40)*.

## 2.3 Monte Carlo simulations

To illustrate that our Dopplergram observations are compatible with and can only be explained by predominance of twist, we perform Monte Carlo simulations. From an observational point of view, we find in a typical field-of-view at the limb:

1. a predominance of redshifted and blueshifted features that are roughly parallel to one another, although not always with the small-scale red/blue pattern that is typical of twist. In other words, we often find seemingly isolated blueshifted or redshifted features that all have the same average orientation but that are not necessarily part of the same individual structure.

2. outward propagation of Doppler-signals at very high phase speed,

3. outward apparent mass motion.

We perform a Monte Carlo simulation of spicules randomly occurring in our synthetic field of view with spicule properties inspired by observations:

- lifetime from a Gaussian distribution around 140s (width of 30s)

- diameters from Gaussian around 400 km (width of 100 km)

- upward velocities from Gaussian around 60 km/s (width of 10 km/s)

- torsional amplitude from Gaussian around 30 km/s (width of 5 km/s)

- Alfvenic propagation with speeds from Gaussian around 200 km/s (width of 30 km/s)

- Alfvenic wave periods from uniform distribution from 120 to 230 s

- random phase of the Alfvenic waves

We perform two simulations:

1. All spicules are vertical, with no non-twisting contribution to the line-of-sight velocity (MC1)

2. Spicule orientation towards or away the observer is chosen from a Gaussian around 0 degrees from vertical (with width of 30 degrees), ensuring that for some spicules the field-aligned flows will contribute to the line-of-sight velocity (MC2)

The resulting synthetic Dopplergrams at 30 km/s are shown in movie S20. We find that whereas MC1 shows the formation and rapid disappearance of nicely twisted features (red/blue pairs), MC2 shows a variety of uniquely redshifted features and uniquely blueshifted features, as well as some twisted features. The synthetic Dopplergrams of MC2 show striking similarity with the observed behavior listed at the top of this section. The properties of MC2 are also the most realistic in terms of what occurs in the solar atmosphere: we expect a variety of viewing angles based on the varying magnetic field direction in the chromosphere and transition region. Our simulations thus show that the observations are fully compatible with the presence of ubiquitous twist, and that our observed Dopplergrams most likely underestimate the presence of twist.

We note that the Monte Carlo simulations did not include the presence of swaying motions, but of course these would have an effect that is similar to that of the field-aligned flows on features inclined toward or away from the observer. Except that swaying motions will affect the line-of-sight velocities even for a perpendicular viewing angle.

Finally, we have assumed the presence of Alfvén waves even if the observations typically only show one wave packet propagating. This use of "waves" for single wave packets is justified when considering the driving force of these disturbances which is magnetic in nature, and the propagation speed (which is of order the Alfvén speed).The ubiquity of twisting motions across open and closed field regions suggest that these motions are more likely being driven from the low solar atmosphere.

## 2.4 Incompatibility with alternative interpretations

In principle, the presence of uniquely redshifted and uniquely blueshifted features in close proximity of each other could be caused by upflowing plasma that occurs on structures that are oriented towards (blue) the observer or away (red) from the observer. However, in such a scenario the blue/red features will only be parallel when projected onto the plane of the sky if the line-of-sight vector between observer and observed structure is within the plane containing both linear features. In reality one would expect that the observer most often observes the plane containing both features at an angle (i.e, from the side), which would lead to blue- and redshifted features that are not parallel when projected onto the plane of the sky. Since we do not see such features, this scenario is unlikely to be occurring.

Another alternative interpretation of red/blue features in close proximity is a scenario in which two features are parallel to one another (thus avoiding the issue above), but one carries a field-aligned upflow, and the other a field-aligned downflow. This is unlikely for several reasons. First, the observations indicate that the red/blue features have their optimal visibility when the line-of-sight is more perpendicular to the linear extent of the features (i.e., the magnetic field direction). This would not occur if field-aligned flows were responsible for the red/blue patterns since the field-aligned component would be minimalized for more perpendicular viewing angles. Secondly, the SST observations indicate mostly outward propagation from the footpoints which is very difficult to explain for field-aligned downflows (i.e., flows towards the footpoints).

A final possible interpretation is one in which shear flows instead of rotational flows are responsible for the red/blue patterns. This can also be excluded. If shear flows were responsibe for the line-of-sight red and blueshifts, one would also expect to see, under most viewing angle conditions, motions in the plane of the sky. Such transverse motions are indeed observed in RBEs *(26,40)*, however their average amplitude is of order 3-8 km/s. This is smaller than the torsional amplitudes we infer here (10-30 km/s). So if the "torsional" velocities were in fact caused by shear flows, they should lead to more vigorous swaying amplitudes than have been measured, since statistically speaking these should be of the same order of magnitude as the line-of-sight flows. This renders it unlikely that shear flows are the dominant cause of the red/blue patterns.

## 2.5 Counterparts in SDO/AIA images

It is not clear how often the associated heating events reach coronal temperatures. Analysis of coordinated observations with IRIS and SDO/AIA indicate that some of these events are visible in several of the AIA channels both in active regions and in quiet Sun (Movie S7), in particular the 304Å and 171Å and occasionally 193Å channels. These are dominated by, respectively, He II 304 Å (~100,000 K), Fe IX/X (~1 MK) and Fe XII (~1.5 MK). Note that the latter "coronal" channels also include contributions from transition region channels (O IV, O V, O VI lines) so it is not yet clear whether and how much plasma is heated to coronal temperatures in association with the heating to transition region temperatures. The AIA counterparts are more evident in the active region observations, suggesting that the densities reached are higher in twisted features that originate in active region plage.

**2.6 Width of twisting features in Dopplergrams**

Some of the features we observe in Dopplergrams appear to be wider than some of the historically reported diameters of spicules (which are of order a few hundred kilometer, see, e.g., (20)). This can be explained by taking into account two issues:

1. Spicules often show collective behavior over up to several arcseconds, i.e., the diameter of individual threads (often reported as the spicule diameter) is not necessarily that of the "collective behavior of a spicule". This was shown for active region spicules *(41)*, but has now also been very clearly demonstrated using Ca II H (Hinode) and Mg II k (IRIS) coordinated observations *(42)*. In other words, the observed and reported spicule diameters are often those of individual threads that take part in a larger scale event.

2. The swaying motions often have larger cross-field coherence than a typical spicules since the swaying motions are related to the a pervasive field of swaying flux tubes, as evidenced in numerical simulations *(2)*. This may lead to the visual impression that "spicules" are wider in Dopplergrams than individual "intensity spicules". Combined with the complex visibility of twist in Figure S2, this can lead to sometimes wider structuring in Dopplergrams.

## 3. Figures

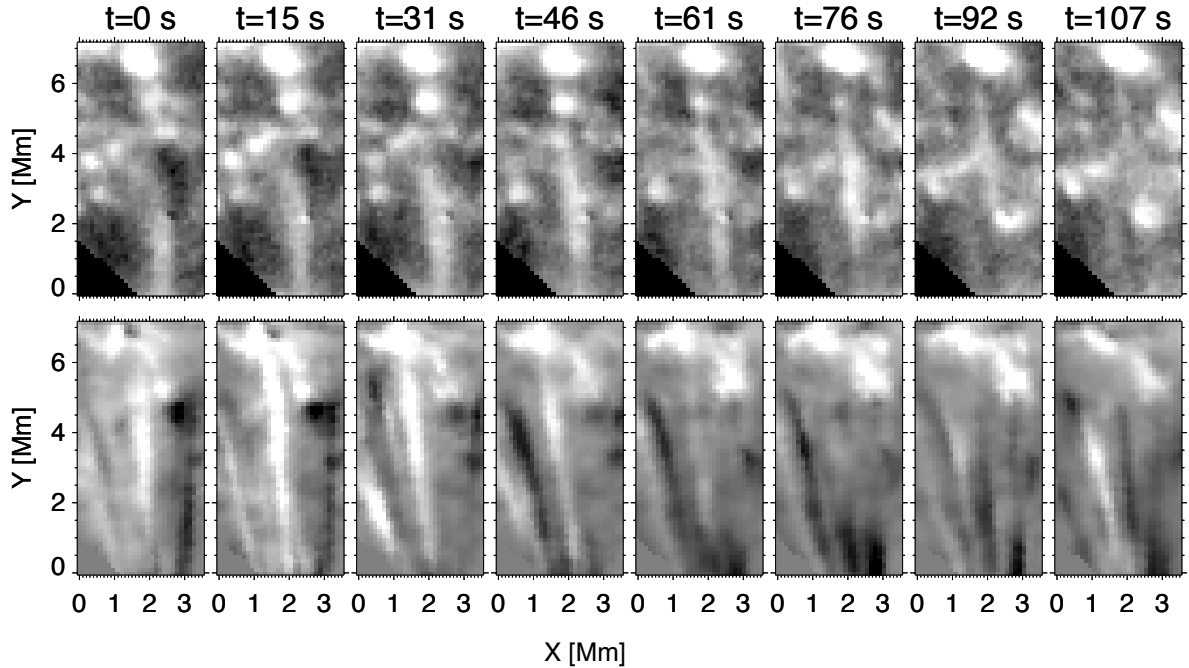

Figure S3: Companion to Figure 3 in main text for a different event in a different region and date (22-Sep-2013). Temporal evolution of twist and associated heating. SST Hα Dopplergrams at ±36 km/s (bottom row) show how quickly chromospheric twist propagates along elongated features on timescales of less than 1 minute. Several of these twisted features are associated with transition region signals (upper row) as observed with the IRIS slit-jaw images that are dominated by C II lines (~30,000 K). Movies S11-S12 provide further examples of the highly dynamic nature of the twist and associated heating.

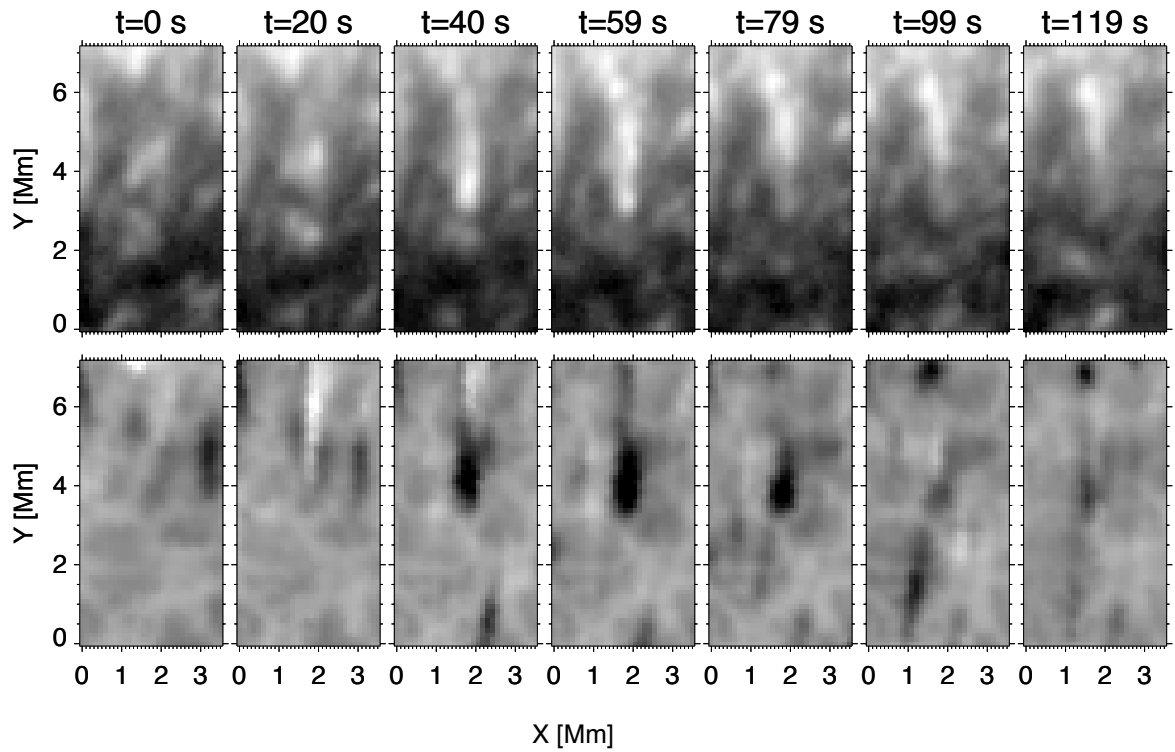

Figure S4: Companion to Figure 3 in main text for a different event in a different region and date (13-Sep-2013). Temporal evolution of twist and associated heating. SST Hα Dopplergrams at ±46 km/s (bottom row) show how quickly chromospheric twist propagates along elongated features on timescales of less than 1 minute. Several of these twisted features are associated with transition region signals (upper row) as observed with the IRIS slit-jaw images that are dominated by Si IV lines (~80,000 K). Movies S13-S14 provide further examples of the highly dynamic nature of the twist and associated heating.

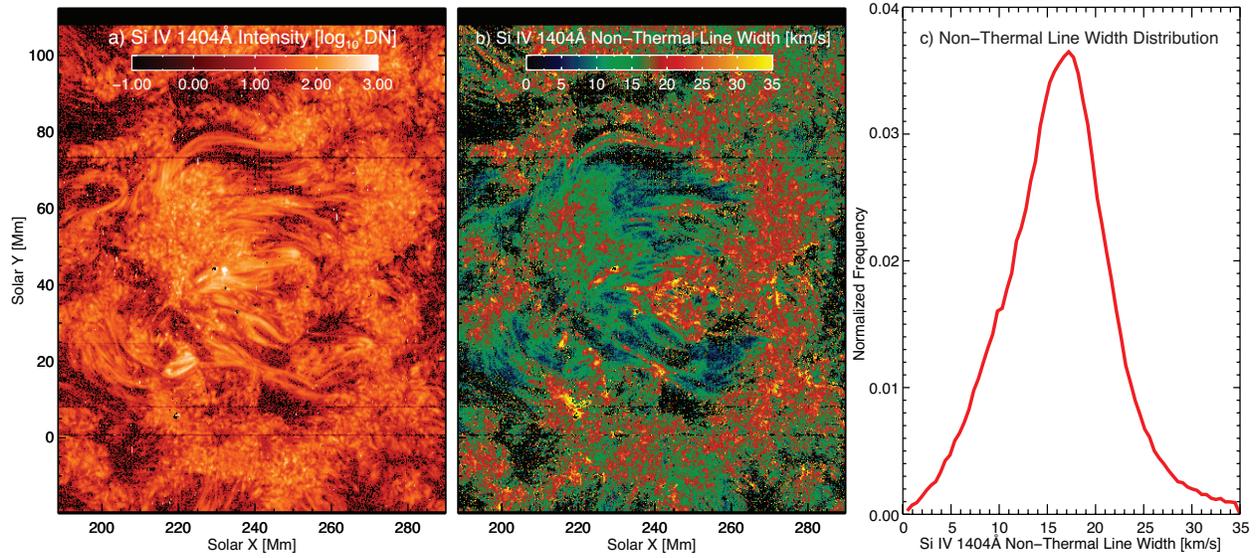

Figure S5: Prevalence of non-thermal broadening in an active region as observed with IRIS Si IV 1404 Å spectra on 27-Sep-2013. Panel a shows the logarithm of the peak intensity of the line, whereas panel b shows the non-thermal line broadening (i.e., in excess of thermal and instrumental broadening) in km/s. Panel c shows a histogram of non-thermal line width which peaks around 17 km/s for this active region with a significant tail out to 35 km/s. It is likely that in some of the features (especially those that are loop-like and appear to be heavily inclined from the local vertical), this broadening is caused by unresolved twist.

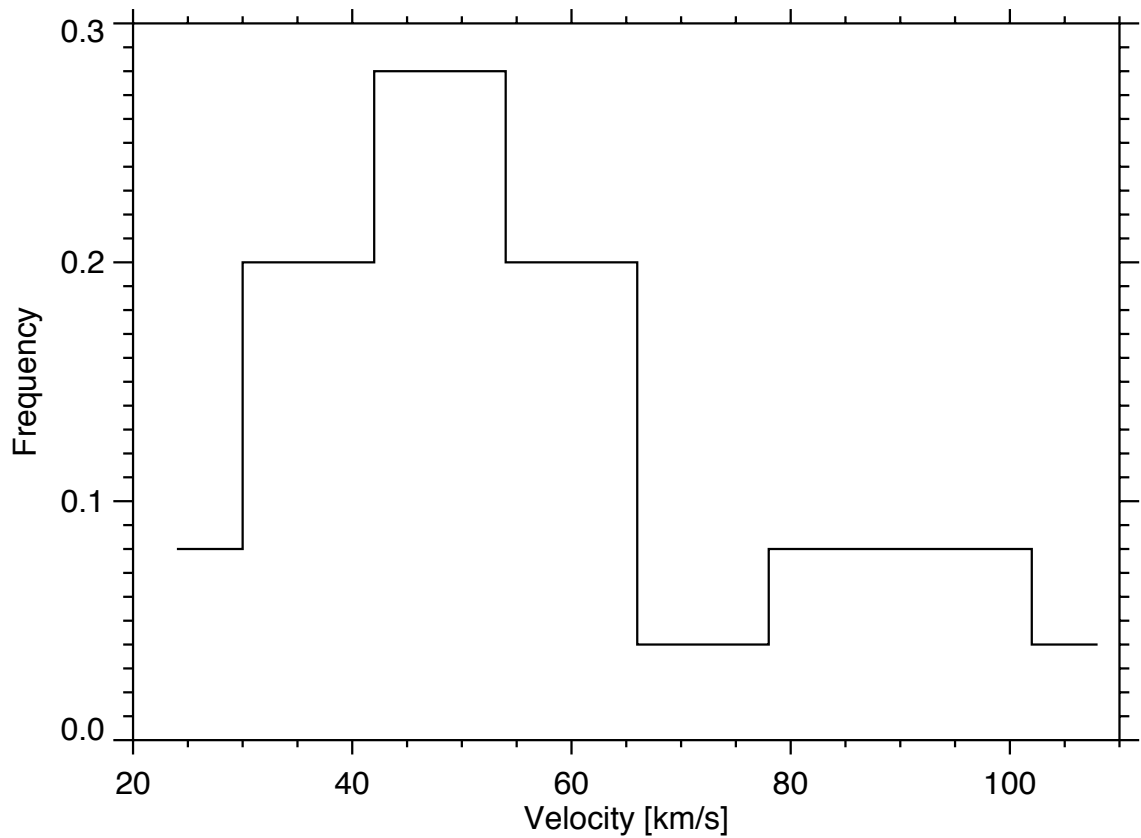

Figure S6: Measurements of apparent propagation of features in Dopplergrams at ±25 km/s using data from a Quiet Sun region at the limb near the North Pole as imaged at very high cadence (1.33s) on 4-Jul-2013 with SST/CRISP in the Ca II 8542Å line (see also Fig. 4 and Movies S15-S19). We measured the apparent propagation speed in the plane of the sky of 25 different black/white (blue- and red-shifted features) features. This is a lower limit to the Alfven speed since the propagation is projected onto the disk. In addition, if these propagating twist features are partially standing waves, the measured propagation could be related to the motion of the nodes of the wave, and thus be a lower limit to the Alfven speed. When focusing on only the blue or red wing, one can often see features appear at much higher apparent phase speeds, compatible with the presence of partially standing waves (see also (43)). The measured values are of the same order as those observed by Kuridze et al. *(45)* who observed kink mode waves.

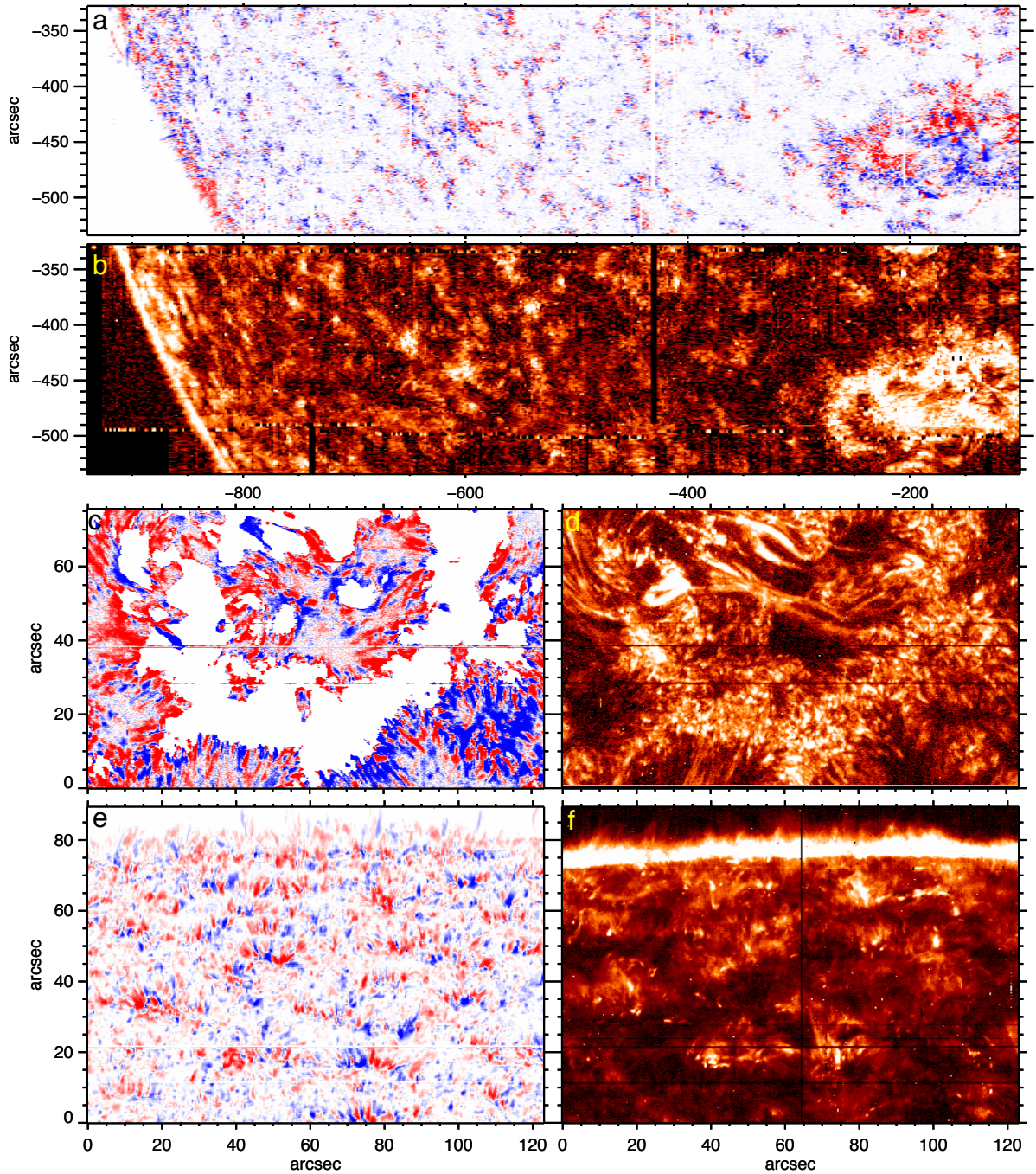

Figure S7: Ubiquity of twist in quiet Sun and active regions. These panels show the same region as that depicted in Figure 1 but with larger field-of-view for the bottom two rows. Dopplergrams (panels a, c, e) formed by subtracting blue- and red-shifted signals in the wings of the chromospheric (~10,000K) Mg II h 2803Å line (at 30 km/s from line center) show a multitude of elongated features in which strongly red- and blueshifted features are parallel and adjacent to each other. These features illustrate that twist is predominant: at the solar limb associated in so-called spicules (a, e), but also in active regions (panel c, and around -250",-475" in panel a), both regions where the line-of-sight is likely more perpendicular to the local magnetic field. Twist is often associated with significant brightening in the transition region (80,000 K) as illustrated with the Si IV 1403Å integrated brightness map in panels b, d and f.

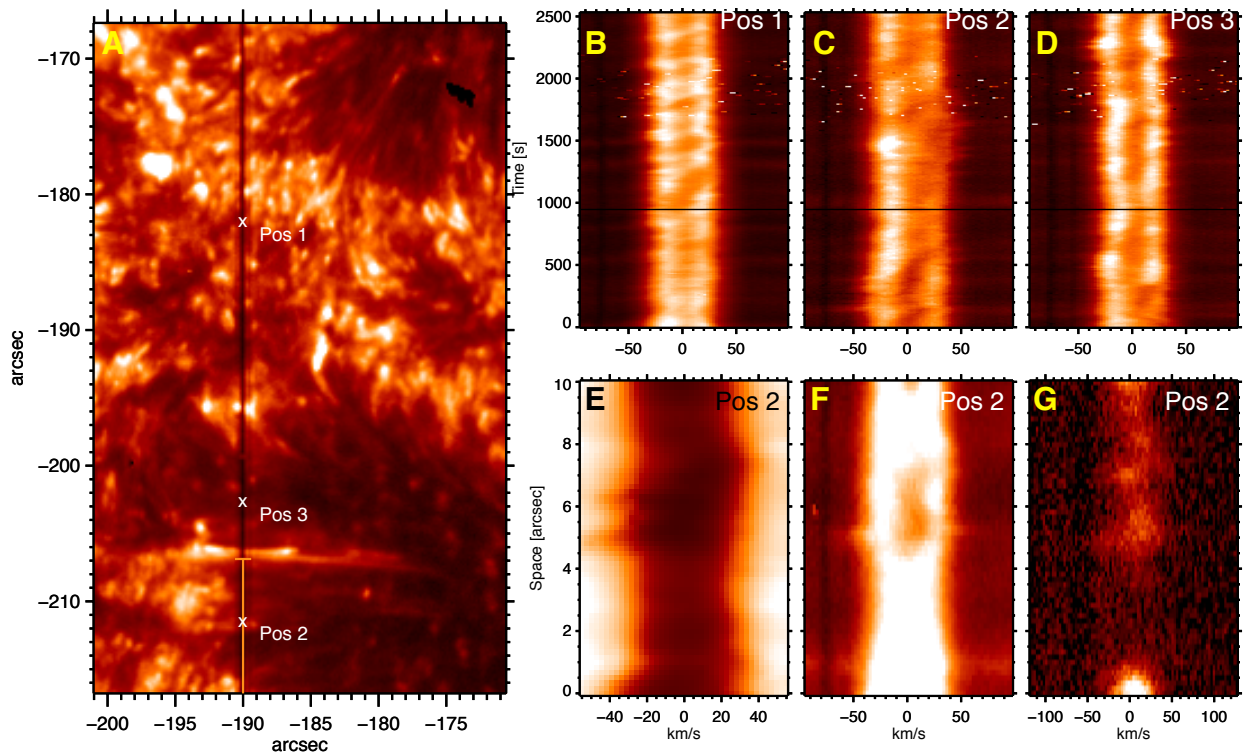

Figure S8: Spatio-temporal properties of twist. Similar to Figure 2 but for another example (Pos 2). Rapidly evolving twisting motions are apparent as short-lived, bright features in the blue and red wings (e.g., around ±50 km/s) of the chromospheric Mg II h 2803Å spectral line (panels C, D) in regions of inclined field (positions 2, 3 shown in panel A, Si IV 1403Å slit-jaw image). These motions contrast with the acoustic shock-dominated spectral profiles (panel B) in position 1 in plage (where field and line-of-sight are more aligned, preventing visibility of twist) that evolve on time-scales of several minutes *(25)*. The spatial pattern of the bright Si IV feature around -190",-212" (position 2 in panel A) is associated with short-lived twisting motions that are visible in Si IV (80,000K, panel G), Mg II h (10,000K, panel F) and faintly in H-alpha (<10,000K, panel E, see movies S5-S7). Velocities of order 30 km/s are reached in this twisting feature. Typical velocities are lower (10-30 km/s) with visibility in these various passbands variable, Mg II h showing excellent visibility in most cases.

## 4. Movies

**Movie S1**

Spectral scan through the Mg II h 2803Å line for the IRIS full disk mosaic taken on 13-Oct-2013. The spectral scan effectively constitutes a scan in height and velocity since the core of the line is formed higher in the chromosphere, whereas the wings are formed either in the photosphere or in Doppler-shifted chromospheric plasma. To improve visibility of twisting motions, we have removed the small Doppler shift caused by solar rotation. Because of the slight uncertainty of order 5" in pointings, some areas of the Sun were not covered by the full-disk mosaic. These regions do not contain any valid data and are visible as vertical stripes that are black in intensity maps, and bright white in Dopplergrams (Fig 1a-1b, Movies S1-S2).

**Movie S2**

Dopplergrams in Mg II h 2803Å line for the IRIS full disk mosaic taken on 13-Oct-2013 and for various offset velocities. Close to the line core this scan mostly reveals differences in line formation of this optically thick line. In the wings the red/blue patterns are indicative of velocities along the line-of-sight between the gas and the observer (IRIS). To improve visibility of twisting motions, we have removed the small Doppler shift caused by solar rotation. Because of the slight uncertainty of order 5" in pointings, some areas of the Sun were not covered by the full-disk mosaic. These regions do not contain any valid data and are visible as vertical stripes that are black in intensity maps, and bright white in Dopplergrams (Fig 1a-1b, Movies S1-S2). Note that the Dopplergram has a slightly different scaling from Fig. 1A.

**Movie S3**

Spectral scan through the Mg II h 2803Å line for the north pole scan taken on 9-Oct-2013. In this case there was no polar coronal hole, so this region is representative for quiet Sun. The spectral scan effectively constitutes a scan in height and velocity since the core of the line is formed higher in the chromosphere, whereas the wings are formed either in the photosphere or in Doppler-shifted chromospheric plasma.

**Movie S4**

Dopplergrams in Mg II h 2803Å line for the north pole scan taken on 9-Oct-2013 and for various offset velocities. Close to the line core this scan mostly reveals differences in line formation of this optically thick line. In the wings the red/blue patterns are indicative of velocities along the line-of-sight between the gas and the observer (IRIS).

**Movie S5**

Overview of AR 11838 on 10-Sep-2013 (also shown in Figure 2) with SST Hα Dopplergrams (at ±36 km/s), IRIS SJI 1400, IRIS SJI 1330 and SDO/AIA 304Å passbands. Also shown are spectra in C II 1334Å, Mg II h 2803Å, Si IV 1403Å and Si IV

1394Å. The twisting event shown in Figure 2 occurs around y=-205". Various other twist events are seen to occur along the slit with clear association of chromospheric Doppler signals to transition region intensities. Since this is a 2-step raster the SJI are taken at different slit positions. The spectra are shown for the eastern position of the raster (red line).

**Movie S6**

Zoom-in of event shown in Figure 2 in AR 11838 on 10-Sep-2013 (also shown in Figure 2). Similar panels as movie S5.

**Movie S7**

Thermal evolution of AR 11838 on 10-Sep-2013 (also shown in Figure 2) with SST Hα Dopplergrams (at ±27 and ±41 km/s, ~10,000 K), IRIS SJI 1330 (~20,000 K), IRIS SJI 1400 (~80,000 K), and SDO/AIA 304Å (0.1 MK), 171 Å (0.8 MK), 193 Å (1.5 MK) and 211 (2 MK) passbands. The twisting event shown in Figure 2 occurs around y=-205" and has clear counterparts in all channels, from 10,000 K to 2 MK.

**Movie S8**

Quiet region network near disk center on 23-Sep-2013 showing the clear association between chromospheric twist/torsional motions (observed in SST Hα Dopplergrams at ±46 km/s, and the blue wing of Hα at -46 km/s, 10,000 K) and transition region emission (IRIS SJI 1330, 20,000 K and SJI 1400, 80,000K). The TR emission is shown with a linear color table. This is the same region as shown in Figure 3.

**Movie S9**

Identical to movie S8, except for a logarithmic color table for the IRIS SJI channels to better deal with the enormous dynamic range of the transition region lines. This movie shows that there is a haze of weak emission that is well correlated with the dynamic twist in the chromosphere. This is the same region as shown in Figure 3.

**Movie S10**

Movie illustrating the temporal evolution of the event in Figure 3. This event occurred on 23-Sep-2013 in the same region that is shown in movies S8 and S9. Note the (white) feature (#1) at t=15s at x=2, y=6 that propagates downward by t=30s (e.g., x=2, y=5), and then fades rapidly by t=44s (barely visible as a lack of signal at x=2, y=6 at t=44s). There is also a second feature (#2) which forms first as a black region at t=0, x=3, y=5.5, first staying as a black feature, and then rapidly showing twist (white/black) from t=59s to t=104s between x=2-3 and y=2-5.5.

**Movie S11**

Network region in a coronal hole on 22-Sep-2013 showing the clear association between chromospheric twist/torsional motions (observed in SST Hα Dopplergrams at ±46 km/s, and the blue wing of Hα at -46 km/s, 10,000 K) and transition region emission (IRIS SJI 1330, 20,000 K and SJI 1400, 80,000K). The TR emission is shown with a logarithmic color table. This is the same region as shown in Figure S3. The black/white pattern around x=520", y=285" in the SST data is an imaging artifact.

**Movie S12**

Movie illustrating the temporal evolution of the event in Figure S3. This event occurred on 22-Sep-2013 in the same region that is shown in movie S11.

**Movie S13**

Quiet region network near disk center on 13-Sep-2013 showing the clear association between chromospheric twist/torsional motions (observed in SST Hα Dopplergrams at ±46 km/s, and the blue wing of Hα at -46 km/s, 10,000 K) and transition region emission (IRIS SJI 1330, 20,000 K and SJI 1400, 80,000K). The TR emission is shown with a logarithmic color table. This is the same region as shown in Figure S4.

**Movie S14**

Movie illustrating the temporal evolution of the event in Figure S4. This event occurred on 13-Sep-2013 in the same region that is shown in movie S13.

**Movie S15**

Quiet Sun region at the limb near the North Pole as imaged at very high cadence (1.33s) on 4-Jul-2013 with SST/CRISP in the Ca II 8542Å line. The Dopplergrams at ±25 km/s and blue and red wing signals at 25 km/s show the mix of various motions described in S2.2 and S2.3, including twist, swaying and field-aligned flows. The very high cadence allows us to track the upward propagation of torsional motions. This movie accompanies Figure 4.

**Movie S16**

Zoom-in of one event in the same region as shown in movie S15. Shows blue/red wing of Ca II 8542Å at 25 km/s as well as Dopplergram at ±25 km/s. The right panel shows the distance/time plot. Accompanies Fig 4a.

**Movie S17**

Same as Movie S16, except for the event shown in Fig 4b.

**Movie S18**

Same as Movie S16, except for the event shown in Fig 4c.

**Movie S19**

Same as Movie S16, except for the event shown in Fig 4d.

**Movie S20**

Synthetic Dopplergram of a Monte Carlo simulation (MC1, see Section S2.3) of solar features showing twist. In the top panel the features are seen from a line-of-sight perpendicular to the long axis of the features, and no non-twisting velocities are present along the line-of-sight, allowing full visibility of twist. In the bottom panel, the features are allowed to be inclined towards or away from the observer, ensuring that some of the field-aligned flows have a component along the line-of-sight. The visibility of twist is significantly reduced or even eliminated.

**Movie S21**

Active region 11856 on 2-Oct-2013 with Mg II k 2796Å and Si IV 1403Å spectra on the left and middle. These spectra were taken along the slit that is shown as a dark line in the SJI 1400 bandpass shown on the right. The multitude of locations that show strong broadening or bi-directional velocities at the limb (where the magnetic field is likely perpendicular to the line-of-sight) indicates that twist occurs throughout the plage regions that the slit crosses.

**Movie S22**

Movie showing the Mg II h Dopplergram (left panel, first frame) and Si IV intensity (right panel) and, for comparison, an "Intensitygram", i.e., an image showing the sum of the blue and red wings of Mg II h (left panel, second frame). Both are shown for the active region shown in Figures 1C and 1D. Blinking these frames shows that the Dopplergrams often show a cross-field gradient of Dopplergram color for what appear to the same structure in the intensity-gram.